\documentclass[aps,prx,superscriptaddress,reprint]{revtex4-1}
\usepackage{amsmath}
\usepackage{amssymb}
\usepackage{graphicx}
\usepackage[colorlinks,urlcolor=blue]{hyperref}
\usepackage{url}
\usepackage{hyperref}
\usepackage{color,xcolor}
\usepackage{epstopdf}
\usepackage{float}
\usepackage{ulem}
\usepackage[percent]{overpic}
\usepackage{siunitx}
\usepackage{todonotes}
\usepackage{lineno}
\usepackage[none]{hyphenat}

\begin{document}

\title{Electronic Structure, Magnetic and Pairing Tendencies of Alternating Single-layer Bilayer Stacking Nickelate La$_5$Ni$_3$O$_{11}$ Under Pressure}
\author{Yang Zhang}
\author{Ling-Fang Lin}
\email{lflin@utk.edu}
\affiliation{Department of Physics and Astronomy, University of Tennessee, Knoxville, Tennessee 37996, USA}
\author{Adriana Moreo}
\affiliation{Department of Physics and Astronomy, University of Tennessee, Knoxville, Tennessee 37996, USA}
\affiliation{Materials Science and Technology Division, Oak Ridge National Laboratory, Oak Ridge, Tennessee 37831, USA}
\author{Satoshi Okamoto}
\affiliation{Materials Science and Technology Division, Oak Ridge National Laboratory, Oak Ridge, Tennessee 37831, USA}
\author{Thomas A. Maier}
\email{maierta@ornl.gov}
\affiliation{Computational Sciences and Engineering Division, Oak Ridge National Laboratory, Oak Ridge, Tennessee 37831, USA}
\author{Elbio Dagotto}
\email{edagotto@utk.edu}
\affiliation{Department of Physics and Astronomy, University of Tennessee, Knoxville, Tennessee 37996, USA}
\affiliation{Materials Science and Technology Division, Oak Ridge National Laboratory, Oak Ridge, Tennessee 37831, USA}

\date{\today}

\begin{abstract}
Nickelates have continued to surprise since their unconventional superconductivity was discovered.
Recently, the layered nickelate La$_5$Ni$_3$O$_{11}$ with hybrid single-layer and bilayer stacking showed superconductivity under high pressure. This compound combines features of La$_2$NiO$_4$ and La$_3$Ni$_2$O$_7$, but its pairing mechanism remains to be understood. Motivated by this finding, here we report a comprehensive theoretical study of this system. Our density functional theory calculations reveal that the undistorted P4/mmm phase without pressure is unstable due to three distortion modes. Increasing pressure suppresses these modes and causes ``charge transfer'' between the single-layer and bilayer sublattices, leading to hole-doping in the single-layer blocks. Our random-phase approximation calculations indicate a leading $d_{x^2-y^2}$-wave pairing state that arises from spin-fluctuation scattering between Fermi surface states mainly originating from the single-layer blocks and additional weaker contributions from the bilayer blocks. These spin-fluctuations could be detected by inelastic neutron scattering as a strong peak at ${\bf q}=(\pi, \pi)$.
Our findings distinguish La$_5$Ni$_3$O$_{11}$ from other nickelate superconductors discovered so far and the high-$T_c$ cuprates. We also discuss both similarities and differences between La$_5$Ni$_3$O$_{11}$ and other hybrid stacking nickelates.
\end{abstract}

\maketitle

\noindent {\bf \\Introduction\\}
Unconventional superconductivity commonly occurs in transition-metal compounds like Cu oxides~\cite{Bednorz1986} and Fe pnictides~\cite{Kamihara2008}.
These materials often feature antiferromagnetic insulators or metals next to superconducting states,
suggesting strong repulsive interactions or magnetic fluctuations as the cause of pairing~\cite{Dagotto:rmp94,Dagotto:rmp13}.
The discovery of superconductivity under pressure in the Ruddlesden-Popper (RP) bilayer (BL) nickelate La$_3$Ni$_2$O$_7$~\cite{Sun:arxiv} with $T_c\sim80$~K and trilayer (TL) La$_4$Ni$_3$O$_{10}$~\cite{Zhu:arxiv11,Li:cpl} with $T_c\sim30$~K opened a new playground for the study of
unconventional superconductivity~\cite{LiuZhe:arxiv,Wang:arxiv9,Dong:arxiv12,Zhang:prb24,Xie:SB,Zhang:4310}.
This field rapidly developed into the newest branch of the high-temperature superconductor family.

The role of high pressure has been clarified. The
BL La$_3$Ni$_2$O$_7$ has Amam or Cmcm structure (No. 63), and the NiO$_6$ octahedra are distorted.
Under hydrostatic pressure, this distortion is suppressed with a first-order structural transition around 14 GPa~\cite{Zhang:arxiv-exp}, leading to a Fmmm phase without tilting distortion, followed by a very broad superconducting region~\cite{Sun:arxiv}.
Very recently, a tetragonal I4/mmm phase was proposed to be the ``real'' structure under high pressure for BL La$_3$Ni$_2$O$_7$ in both theory~\cite{Geisler:qm} and experiment~\cite{Wang:jacs}, instead of the Fmmm structure. Since the distortion from the I4/mmm (No. 139) to
Fmmm (No. 69) is tiny~\cite{Zhang:1313}, there are no fundamental differences among those two phases, providing the same physics~\cite{Sakakibara:prl24}. The initial conclusion of having a superconducting transition in La$_3$Ni$_2$O$_7$ was based on measurements of the resistance using a four-terminal device on a sample with an unknown degree of inhomogeneity, where a sharp transition and a flat stage in resistance was observed by using KBr as the pressure-transmitting medium, as well as a diamagnetic response in the susceptibility. This was interpreted as indication of the two prominent properties of superconductivity, zero resistance and Meissner effect~\cite{Sun:arxiv}. Subsequent experiments confirmed the zero resistance by several independent studies~\cite{Hou:arxiv,Zhang:arxiv-exp,Zhang:jmst}. Recently, the Meissner effect of the superconducting state of La$_3$Ni$_2$O$_7$ was observed by using the $ac$ magnetic susceptibility reporting that the superconducting volume fraction is robust, around $48 \%$~\cite{Li:arxiv24}.

\begin{figure}
\centering
\includegraphics[width=0.46\textwidth]{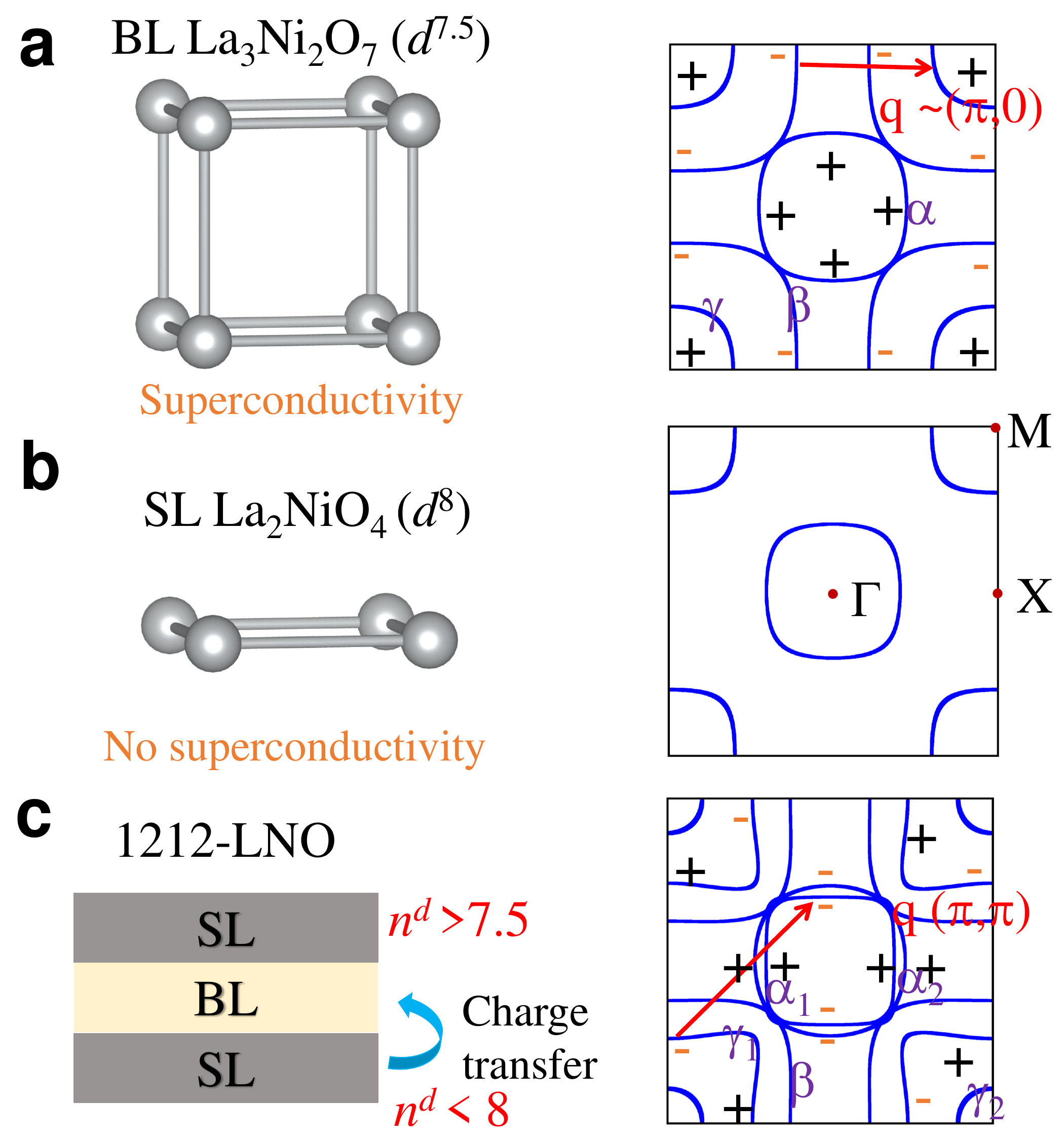}
\caption{{\bf Comparison between RP nickelate La$_{n+1}$Ni$_n$O$_{3n+1}$ and 1212-LNO.}
{\bf a} BL Ni lattice and the sketch of the Fermi surface of La$_3$Ni$_2$O$_7$, which superconducts under pressure.
As shown in the right panel, an $s_{\pm}$-wave channel is favored due to the nesting vector ${\bf q}=(\pi, 0)/(0, \pi)$ connecting the M = $(\pi, \pi)$ centered pockets and portions of the Fermi surface centered at the X = $(\pi, 0)$ [or Y = $(0, \pi)$] points.
{\bf b} SL Ni lattice and the sketch of the Fermi surface of La$_2$NiO$_4$, which does not superconduct at both ambient and high-pressure condition.
{\bf c} SL BL stacking Ni superlattice and the sketch of the Fermi surface of 1212-LNO, which superconducts under pressure.
In contrast to the BL nickelate, a $d_{x^2-y^2}$-wave channel is favored due to the nesting vector ${\bf q}=(\pi, \pi)$ connecting the M = $(\pi, \pi)$ centered sheets and portions of the Fermi surface centered at the $\Gamma$ sheets, as shown in the right panel.}
\label{Big-picture}
\end{figure}

Density functional theory (DFT) calculations have played a crucial role to understand the electronic properties of BL nickelate superconductors.
It was suggested that the Fermi surfaces are made up of Ni $d_{x^2-y^2}$ and $d_{3z^2-r^2}$ orbitals~\cite{Luo:prl23,Zhang:prb23} with two-electron sheets with mixed $e_g$ orbitals. A small hole pocket from the Ni $d_{3z^2-r^2}$ orbital was absent without pressure but appears at high pressure (see Fig.~\ref{Big-picture}~{\bf a}). The angle-resolved photoemission spectroscopy experiments confirmed the two-electron sheets at the Fermi surface without pressure~\cite{Yang:arxiv09}.
Theoretical studies suggested that $s_{\pm}$-wave pairing superconductivity is dominant,
induced by spin fluctuations due to the partial nesting of the Fermi surfaces with wave vectors ($\pi$, 0) or (0,$ \pi$)~\cite{Yang:prb23,Zhang:nc24,Liu:prl23,Liao:prb23,Qu:prl,Lu:prl}.
The $s_{\pm}$-wave pairing channel was believed to be driven by strong
interlayer coupling, but the $d_{x^2-y^2}$ orbital also has robust contributions to the superconducting gap, comparable to the contributions of the $d_{3z^2-r^2}$ orbital~\cite{Tian:prb24,Zhang:prb23-2}. In addition, some theoretical studies, alternatively, suggested the possibility of $d$-wave pairing arising from the superconducting pairing state in the $\beta$ sheet, mainly driven by intralayer coupling~\cite{Lechermann:prb23,Jiang:cpl,Fan:prb24,Xie:nc25,Heier:prb24}.

\begin{figure}
\centering
\includegraphics[width=0.9\columnwidth]{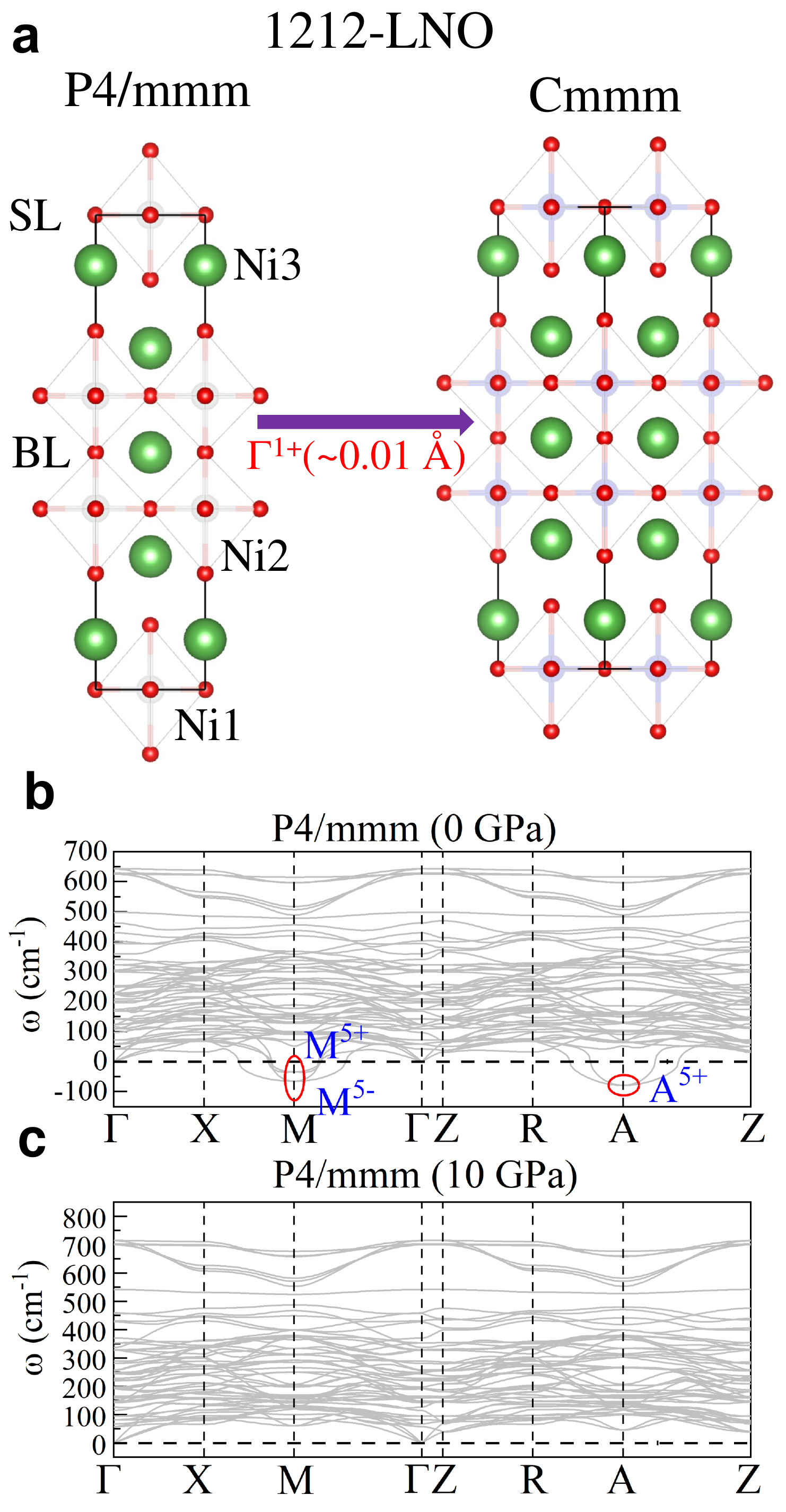}
\caption{ {\bf Crystal structure and phonon spectrum.} {\bf a} Schematic crystal structures of conventional P4/mmm (No. 123) and Cmmm (No. 65) phases of 1212-LNO (green = La; blue = Ni; red = O). All crystal structures were visualized using the VESTA code~\cite{Momma:vesta}.
{\bf b, c} Phonon spectrum of the P4/mmm (No. 123) phase of 1212-LNO at 0 GPa, and at 10 GPa, respectively.
The results at other pressures can be found in Supplementary Note II.
}
\label{crystal}
\end{figure}

Similar to BL La$_3$Ni$_2$O$_7$,  TL La$_4$Ni$_3$O$_{10}$ does not superconduct under ambient pressure.
However, applying pressure suppresses the distortion of the NiO$_6$ octahedra and results in superconducting states over a broad pressure range~\cite{Zhu:arxiv11,Li:cpl}.
By contrast, single-layer (SL) La$_2$NiO$_4$ (see Fig.~\ref{Big-picture}~{\bf b}) does not superconduct under any conditions, whether at ambient or high pressure~\cite{Zhang:jmst}.
The superconductivity in the RP nickelates La$_{n+1}$Ni$_n$O$_{3n+1}$ ($n = 1, 2$, and 3) is thus tuned by the layer thickness $n$.
It is natural to ask (1) is there an optimal $n$? and (2) can the superconductivity appear when $n$ is varied in a single crystal, even including SL La$_2$NiO$_4$?
In an effort to answer these questions,
hybrid alternating stacking nickelates were experimentally synthesized, including
the SL-TL stacking nickelate La$_2$NiO$_4$/La$_4$Ni$_3$O$_{10}$ (1313-LNO)~\cite{Chen:jacs,Puphal:arxiv12,Wang:ic,Abadi:arxiv24}
as well as the SL-BL stacking nickelate La$_2$NiO$_4$/La$_3$Ni$_2$O$_7$ (1212-LNO)~\cite{Shi:arxiv25,Li:prm24}.
Surprisingly, a recent experiment found superconductivity in 1212-LNO under pressure~\cite{Shi:arxiv25}.
In contrast to other RP nickelates, an orthorhombic Cmmm phase without tilting distortion was proposed as the structure of 1212-LNO~\cite{Shi:arxiv25} at 0 GPa,
and the application of pressure stabilizes a tetragonal P4/mmm structure (see Fig.~\ref{crystal}~{\bf a}) around 4.5 GPa.
This exciting discovery triggered the study of superconductivity in the hybrid stacking nickelate superlattices.
Several questions naturally arise: what are the electronic and magnetic properties of this 1212-LNO material?
What is the pairing channel under pressure?
What are the similarities and differences between 1212-LNO and other RP hybrid stacking nickelate superlattices,
as well as other RP nickelates?

To answer these questions, here we systematically investigate the alternating SL and BL hybrid stacking nickelate superlattice 1212-LNO by using DFT and random phase approximation (RPA) calculations. The electronic structure of La$_5$Ni$_3$O$_{11}$ arises from the combination of La$_2$NiO$_4$ and  La$_3$Ni$_2$O$_7$.
Our DFT calculations suggest that P4/mmm and the previously proposed Cmmm phases have nearly identical electronic band structure, because the distortion from P4/mmm to Cmmm symmetry is tiny ($\sim 0.01$ \AA).
However, our phonon calculation using the P4/mmm phase without pressure showed imaginary frequencies,
indicating that the high-symmetry undistorted P4/mmm phase is unstable.
Imaginary frequencies appear along high symmetry directions, corresponding to three unstable distortion modes ($A^ {5+}$, $M^{5+}$, and $M^{5-}$) of the oxygen octahedra.
By applying pressure, all unstable distortion modes are gradually suppressed, stabilizing the P4/mmm phase at around 12 GPa.

Key differences between the RP nickelates and 1212-LNO and our findings are highlighted in Fig.~\ref{Big-picture}.
First, despite the fact that 1212-LNO consists of SL and BL blocks, we find a ``charge transfer'' effect from the SL to the BL sublattice, resulting in electron-doping of the BL blocks, compared to the BL RP nickelate La$_3$Ni$_2$O$_7$.
Second, using RPA calculations we find a leading $d_{x^2-y^2}$-wave pairing state with dominant contributions from the SL blocks and additional weaker contributions from the outer $\alpha_2$ sheet from the BL blocks.
Furthermore,  our spin susceptibility calculations show a strong peak at ${\bf q}=(\pi, \pi)$ and another weaker peak at ${\bf q}=(\pi, 0)$.
The former corresponds to the nesting vector $q^M=(\pi,\pi)$ connecting the $\alpha_1$ and the $\gamma_1$ sheets.
These magnetic fluctuations give rise to the $d_{x^2-y^2}$-wave pairing state. The latter corresponds to the nesting vector $q^X=(\pi,0)$, connecting the $\gamma_1$ sheets.
Our analysis unveils similarities and differences between 1212-LNO and other nickelates, including 1313-LNO and La$_3$Ni$_2$O$_7$/La$_4$Ni$_3$O$_{10}$ (2323-LNO) stacking nickelate superlattice structures. Our results also suggest the interface could play an important role in the nickelate superlattice structures.

\noindent {\bf \\Results\\}
\noindent {\small \bf \\Structural instability\\}
We begin by briefly reviewing the structural properties of nickelate superconductors. Single-crystal X-ray diffraction experiments suggest the space group of this hybrid RP 1212-LNO is Cmmm (No. 65) at ambient conditions and the space group transforms to P4/mmm (No. 123) at high pressure, where the out-of-plane Ni-O-Ni angle between the NiO$_6$ octahedrons is flat with a 180$^{\circ}$ bond in both phases~\cite{Shi:arxiv25}, as displayed Fig.~\ref{crystal}~{\bf a}, which is different from the 168$^{\circ}$ bond in the Amam phase of BL La$_3$Ni$_2$O$_7$~\cite{Sun:arxiv}. We calculated the energy for the optimized Cmmm and P4/mmm phases of 1212-LNO. These phases have nearly degenerate energy at 0 GPa due to a quite small distortion amplitude of $\Gamma^{1+}$ ($\sim 0.01$ \AA). Thus, the electronic structure of the Cmmm phase is almost identical to that of the P4/mmm structure [see Supplementary Note I]. This is quite similar to the previous study of 1313-LNO, where the lower Cmmm and higher I4/mmm symmetries both without tilting of the NiO$_6$ octahedrons give the same physics due to the tiny distortion that exists between the I4/mmm and Cmmm phases of 1313-LNO~\cite{Zhang:1313}.

\begin{figure*}
\centering
\includegraphics[width=0.92\textwidth]{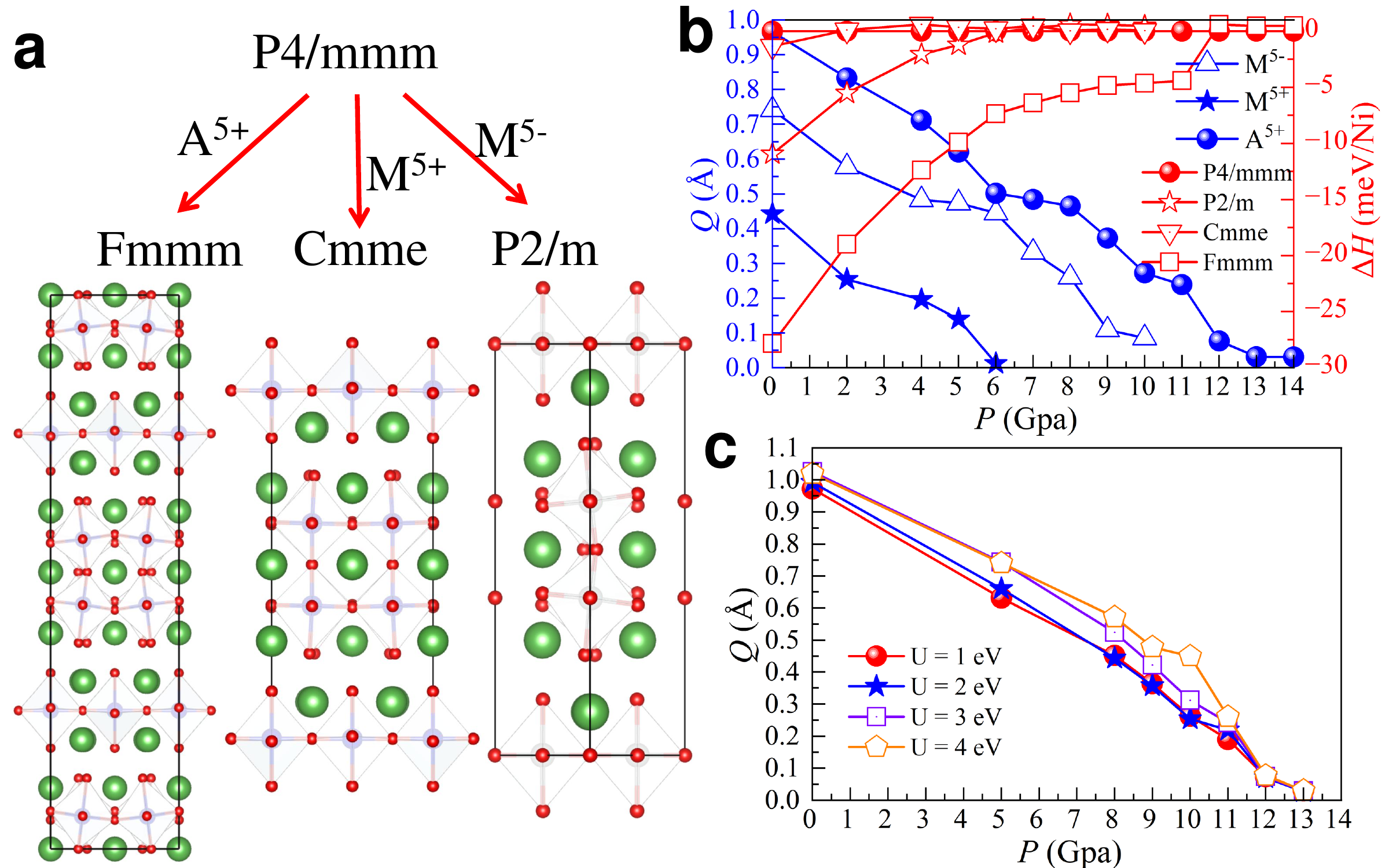}
\caption{ {\bf Group theory analysis, distortion amplitudes, and phase transition.} {\bf a} The group theory analysis for 1212-LNO is based on the unstable phonon modes for the P4/mmm phase at 0 GPa and the schematic crystal structures led by those modes. All crystal structures were visualized using the VESTA code~\cite{Momma:vesta}. {\bf b} The distortion amplitude of different distortion modes and calculated enthalpies (H = E + PV) of different structures, as a function of pressure. {\bf c} The enthalpy of the P4/mmm phase was taken as the reference energy. {\bf c} The $A^{5+}$ distortion amplitudes vs. pressure for different $U_{\rm eff}$.}
\label{Phase-transition}
\end{figure*}

To understand the structural instability of the alternating SL-BL stacking 1212-LNO, we begin by analyzing the phonon spectrum of the high-symmetry P4/mmm phase (No.123) without any tilting at 0 GPa, by using the density functional perturbation theory approach~\cite{Baroni:Prl,Gonze:Pra1,Gonze:Pra2} analyzed by the PHONONPY software~\cite{Chaput:prb,Togo:sm}. As shown in Fig.~\ref{crystal}~{\bf b}, the phonon dispersion spectrum has imaginary frequencies at the high symmetry M and A points at 0 GPa for the P4/mmm phase of 1212-LNO. Furthermore, the phonon dispersion spectrum of Cmmm phase of 1212-LNO is also not stable [see Supplementary Note I], suggesting that Cmmm structure is not the solution of 1212-LNO without pressure. By applying pressure  to the P4/mmm phase of 1212-LNO, imaginary frequencies are suppressed and completely eliminated at around 9 GPa  up to 40 GPa (the maximum pressure we studied). As an example, Fig.~\ref{crystal}~{\bf c} shows the phonon spectrum at 10 GPa. This feature is similar to nickelate superconductors studied previously~\cite{Zhang:nc24,Zhang:arxiv24}, where the untilted high-symmetry structure is not stable at ambient pressure, but becomes stable at high pressure.

Based on the group analysis for the imaginary frequencies of the phonon dispersion spectrum by using the AMPLIMODES software~\cite{Chaput:prb,Togo:sm}, three distortion modes were obtained: $A^{5+}$, $M^{5+}$, and $M^{5-}$. Figure~\ref{Phase-transition}~{\bf a} shows that those distortion modes would lead to low-symmetry Fmmm, Cmme, and P2/m phases from the high-symmetry P4/mmm phase, where the oxygen NiO$_6$ octahedron is distorted in the plane, resulting in a tilting Ni-O-Ni bond. By fully optimizing the crystal structures of different structures, we found that the Fmmm phase of 1212-LNO has the lowest energy among those candidates at ambient pressure. The Fmmm phase with octahedral tilting distortion was also proposed as the possible structure in recent experimental work for 1313-LNO at ambient pressure~\cite{Puphal:arxiv12}, where DFT work also suggested that the Fmmm phase has lower energy~\cite{Zhang:1313}. By applying pressure, the distortion amplitude of different distortion modes smoothly decreases, as well as the difference in enthalpy, between different structural phases. As displayed in Fig.~\ref{Phase-transition}~{\bf b}. At the $A^{5+}$ mode, corresponding to the largest imaginary frequency for the P4/mmm phase at 0 GPa, was fully suppressed around 12 GPa. Moreover, we also tested the effect of the onsite Coulomb interactions by using DFT+$U_{\rm eff}$ $(U_{\rm eff} = U -J)$ with the Dudarev formulation~\cite{Dudarev:prb}. The critical pressure was also obtained around 12 GPa for different $U_{\rm eff}$'s as shown in Fig.~\ref{Phase-transition}~{\bf c}, suggesting the critical pressure is independent of  $U_{\rm eff}$.

\begin{figure*}
\centering
\includegraphics[width=2\columnwidth]{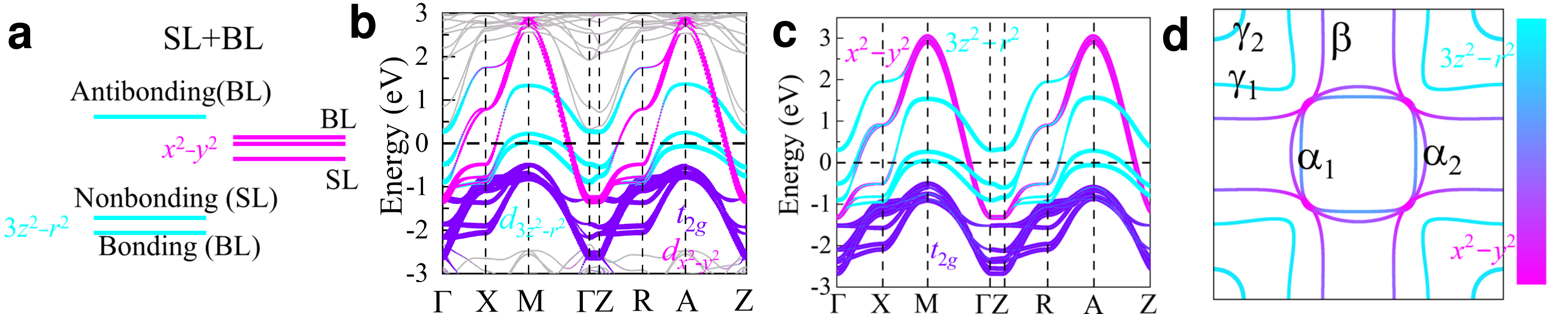}
\caption{{\bf Electronic structure of 1212-LNO in the P4/mmm phase.}
{\bf a} Alignment of $e_g$ levels. Light blue (pink) horizontal lines represent $d_{3z^2-r^2}$ ($d_{x^2-y^2}$) states.
$d_{3z^2-r^2}$ orbitals from the BL sublattice show bonding-antibonding splitting, while $d_{3z^2-r^2}$ orbital coming from the SL sublattice does not show such splitting.
{\bf b} The projected DFT band structure of the P4/mmm phase of 1212-LNO at 20 GPa.
The coordinates of the high-symmetry points in the Brillouin zone are $\Gamma$ = (0, 0, 0), X = (0, 0.5, 0), M = (0.5, 0.5, 0), Z = (0, 0, 0.5), R = (0, 0.5, 0.5) and A = (0.5, 0.5, 0.5). {\bf c} Tight-binding band structure and {\bf d} Fermi surface for the P4/mmm phase of 1212-LNO at 20 GPa.
The input file of the hopping matrices and crystal-field splittings can be found in a separate attachment in the Supplemental Materials.
The tight-binding results at other pressures can be found in Supplementary Note III.}
\label{TB-20}
\end{figure*}

\noindent {\small \bf \\Electronic properties\\}
We now turn to a detailed discussion of the electronic properties of 1212-LNO. In the BL La$_3$Ni$_2$O$_7$ system, due to the ``dimer'' physics induced by its bilayer geometry and robust hopping perpendicular to the planes~\cite{Zhang:prb23}, the $d_{3z^2-r^2}$ orbital splits into bonding and antibonding states, while the $d_{x^2-y^2}$ orbital remains decoupled among the planes. In the SL La$_2$NiO$_4$ system, the $d_{3z^2-r^2}$ orbital does not have a bonding-antibonding splitting.
In this SL-BL 1212-LNO, the energy splitting of the $e_g$ orbitals is shown in Fig.~\ref{TB-20}~{\bf a}, resulting in a layer-selective behavior.

Let us focus on the P4/mmm phase of 1212-LNO at high pressure. First, using first-principles DFT~\cite{Kresse:Prb,Kresse:Prb96,Blochl:Prb,Perdew:Prl}, we calculated the electronic structure of 1212-LNO at 20 GPa, as displayed in Fig.~\ref{TB-20}~{\bf b}, which looks like the combination between SL La$_2$NiO$_4$ and BL La$_3$Ni$_2$O$_7$, where the states near the Fermi level mainly are contributed by the Ni $e_g$ orbitals. Furthermore, the DFT band structure clearly reflects the band alignment discussed here.

Based on the crystal-field splitting and hoppings obtained for 1212-LNO by using the maximally localized Wannier functions (MLWFs) method to map DFT bands on the WANNIER90 package code~\cite{Mostofi:cpc}, we constructed a fifteen-band ${3d}$-orbital tight-binding model containing the 5 $d$-orbitals of the SL
and the 2$\times$5 $d$-orbitals of the BL lattice neglecting longer-range hoppings that connect the SL and BL sublattices. Our model Hamiltonian is given by
\begin{eqnarray}\label{eq:H}
H_k = \sum_{\substack{i \gamma\gamma' \\\vec{\alpha}\sigma}}t_{\gamma\gamma'}^{\vec{\alpha}}
(c^{\dagger}_{i \gamma\sigma}c^{\phantom\dagger}_{i+\vec{\alpha} \gamma'\sigma}+H.c.)+ \sum_{i \gamma\sigma} \Delta_{\gamma} n_{i\gamma\sigma}\,.
\label{Hk}
\end{eqnarray}
Here, the first term represents the hopping of an electron from orbital $\gamma'$ at site $i+\vec{\alpha}$ to orbital $\gamma$ at site $i$.
$c^{\dagger}_{i\gamma\sigma}$ ($c^{\phantom\dagger}_{i\gamma\sigma}$) is the creation (annihilation) operator of an electron at site $i$, orbital $\gamma$ with spin $\sigma$.
$\Delta_{\gamma}$ represents the crystal field of orbital $\gamma$
with $n_{i \gamma \sigma }=c_{i \gamma \sigma}^\dag c_{i \gamma \sigma}$.
The vectors $\vec{\alpha}$ are along the three directions, defining different hopping neighbors (the entire hopping file is available in a separate attachment in the Supplemental Materials). Considering the stoichiometry of 1212-LNO, the electron filling of the $d$ bands is taken as $N = 23$.

We plot the tight-binding band structure of 1212-LNO in Fig.~\ref{TB-20}{\bf c}, by using the parameters obtained at 20 GPa. Another point to note is that, because of the large in-plane hopping amplitude, the $d_{x^2-y^2}$ bands have a wide band width of $\sim 4$~eV. Thus, the $d_{x^2-y^2}$ states have a small density of states near the Fermi level and, therefore, do not contribute significantly to the pairing instability. In addition, the BL $d_{3z^2-r^2}$ bonding state crosses the Fermi level at 20 GPa, leading to a small hole pocket $\gamma_2$ in the Fermi surface, as displayed in Fig.~\ref{TB-20}{\bf d}, where this pocket was believed to be important for the superconductivity in the previous context of the BL La$_3$Ni$_2$O$_7$. Furthermore, the nonbonding state of the $d_{3z^2-r^2}$ orbital from the SL layer crosses the Fermi level, resulting in a hole $\gamma_1$ sheet at the M point (See Fig.~\ref{TB-20}{\bf d}). Compared with the SL La$_2$NiO$_4$ bulk system, the shape of this sheet considerably increases, suggesting a hole-doped scenario in the $\gamma_1$. Similar to the SL-TL stacking 1313-LNO, we also observe a robust ``charge transfer'' between the SL and BL layers, leading to hole-doping and electron-doping in the SL or BL layers, respectively, compared to the stoichiometric carrier concentration of SL La$_2$NiO$_4$ and BL La$_3$Ni$_2$O$_7$.

\noindent {\small \bf \\Pairing symmetry and magnetic correlations\\}
In the previous subsection, we have seen that the electronic state of the SL-BL stacking 1212-LNO is notably different from those in the SL La$_2$NiO$_4$ and the BL La$_3$Ni$_2$O$_7$, even though it can be regarded as a combination of the SL and BL compounds.
Here, we apply many-body RPA calculations to investigate the pairing instability of the multi-orbital Hubbard model, $H=H_k+H_{int}$. $H_k$ is the single-particle part of the Hamiltonian obtained from the P4/mmm phase at 20 GPa, given by Eq.~(\ref{Hk}).
The RPA calculations of the pairing vertex are based on a perturbative weak-coupling expansion with respect to the electron-electron interaction part of the Hamiltonian $H_{int}$.

As detailed in the Methods section, the local Coulomb interaction matrix contains the intra-orbital ($U$), inter-orbital ($U'$), Hund's rule coupling ($J$), and pair-hopping ($J'$) terms~\cite{Kubo2007,Graser2009,Altmeyer2016,Romer2020}.
Both spin and charge susceptibilities enter directly in the pairing interaction for the states on the Fermi surface.
The pairing strength $\lambda_\alpha$ for the pairing channel $\alpha$, and its corresponding pairing gap structure $g_\alpha({\bf k})$, are obtained by solving an eigenvalue
problem, which includes a momentum integral on the Fermi surface involving the irreducible particle-particle vertex $\Gamma({\bf k - k'})$ (for details, see the Method section and Eq.~(\ref{eq:pp})).
The dominant term entering $\Gamma({\bf k-k'})$ is the RPA spin susceptibility tensor
$\chi^s_{\ell_1\ell_2\ell_3\ell_4}({\bf k-k'})$, where $\{\ell_i\}$ denote the fifteen Ni-$d$ orbitals of the 1212-LNO system.
To compute $\chi^s_{\ell_1\ell_2\ell_3\ell_4}({\bf k-k'})$, we have used $U=0.55$ eV, $U'=U/2$, $J=J'=U/4$, and a temperature of $T = 0.02$ eV.
We also note that, by increasing $U$ in our RPA calculations, a spin-density wave instability was obtained at $U \agt 0.6$~eV at $q^M = (\pi,\pi)$  signaled by a divergence of the spin susceptibility due to near-perfect nesting of the SL $\alpha_1$ and $\gamma_1$ sheet Fermi surface states.

\begin{figure}
\centering
\includegraphics[width=0.46\textwidth]{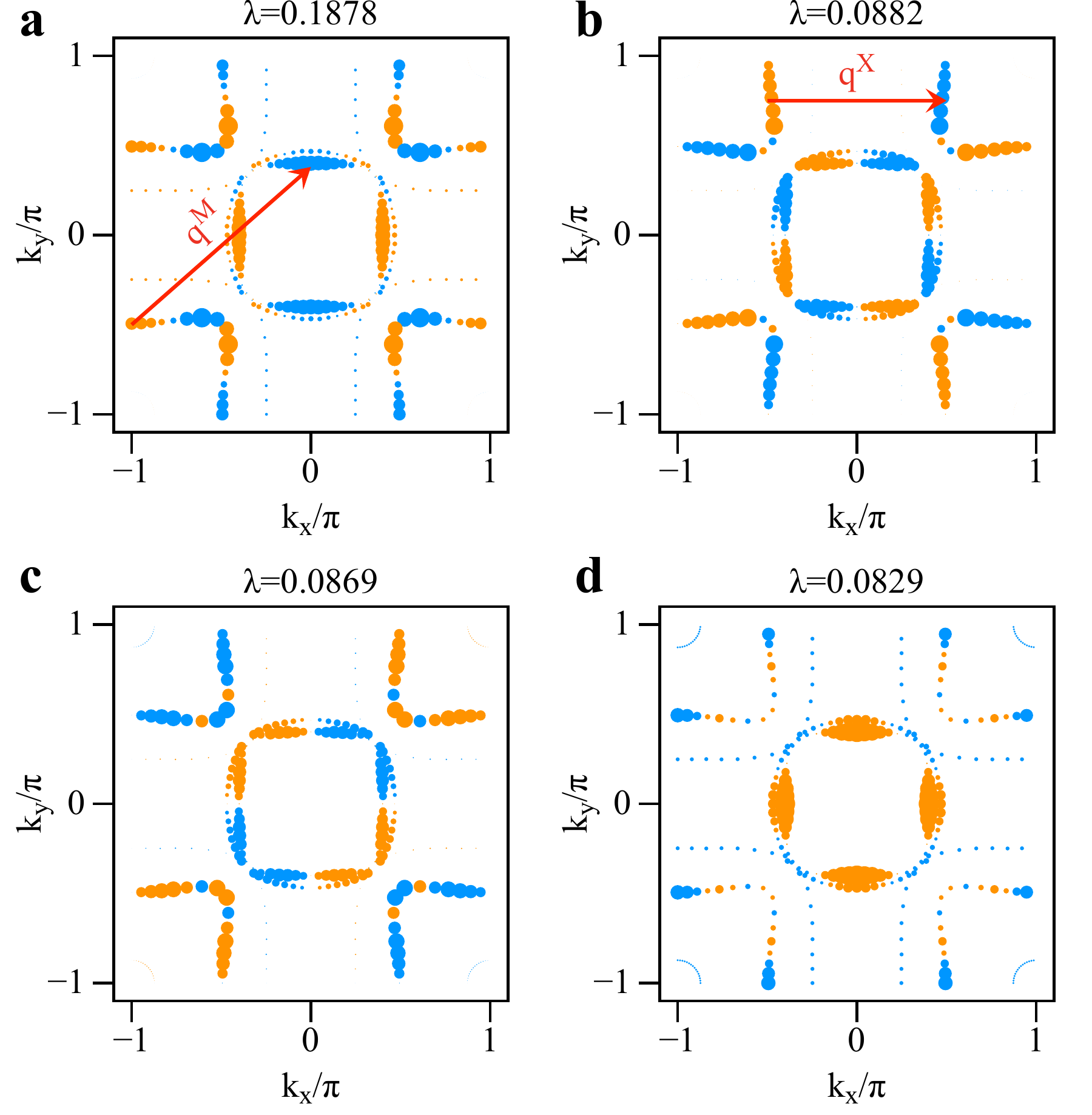}
\caption{ {\bf RPA superconducting gap structure $g({\bf k})$  of 1212-LNO at 20 GPa.} {\bf a} Leading $d_{x^2-y^2}$-wave state with $\lambda=0.1878$;
{\bf b} Subleading $g$-wave state with $\lambda=0.0882$; {\bf c} Subleading $d_{xy}$-wave state with $\lambda=0.0869$; {\bf d} Subleading $s^\pm$-wave state with $\lambda=0.0829$.
The sign of $g_\alpha({\bf k})$ is indicated by the color (orange = positive, blue = negative), and its amplitude by the point sizes. Here we used Coulomb parameters $U=0.55$ eV, $U'=U/2$, and $J=J'=U/4$, and the calculation was performed for a temperature of $T=0.02$ eV.}
\label{RPA}
\end{figure}

By solving the eigenvalue problem of Eq.~(\ref{eq:pp}) within the RPA, we find that the leading pairing instability appears in the $d_{x^2-y^2}$-wave channel with the largest eigenvalue $\lambda = 0.1878$. For this state, the gap is largest on the inner $\alpha_1$ and the $\gamma_2$ sheets and its phase changes sign between these two sheets (see Fig.~\ref{RPA}~{\bf a}). The leading $d_{x^2-y^2}$-wave instability is well separated from the subleading solutions with $g$-wave, $d_{xy}$-wave, and $s^\pm$-wave states which all have nearly degenerate pairing strength $\lambda$, as shown in Figs.~\ref{RPA}~{\bf b-d}.

The leading $d_{x^2-y^2}$-wave instability is well separated from the other subleading gap structures with much smaller eigenvalues. Furthermore, for all pairing states obtained here, pairing is strong mainly for the SL sublattice Fermi surface states, and, to a lesser extent, on the outer $\alpha_2$ sheet that originates from the BL sublattice. In contrast, the $\beta$ sheet and the $\gamma_2$ pocket do not contribute to pairing states in the 1212-LNO case. Compared to the pure BL stacking La$_3$Ni$_2$O$_7$ system (3 electrons for $e_g$ orbitals in two Ni sites), the BL block layers of 1212-LNO are  ``electron-doped'' ($\sim 3.247$ electrons for $e_g$ orbitals in two Ni sites), and the superconducting pairing appears to be strongly suppressed in the BL sublattice of 1212-LNO.

\begin{figure}
\centering
\includegraphics[width=0.46\textwidth]{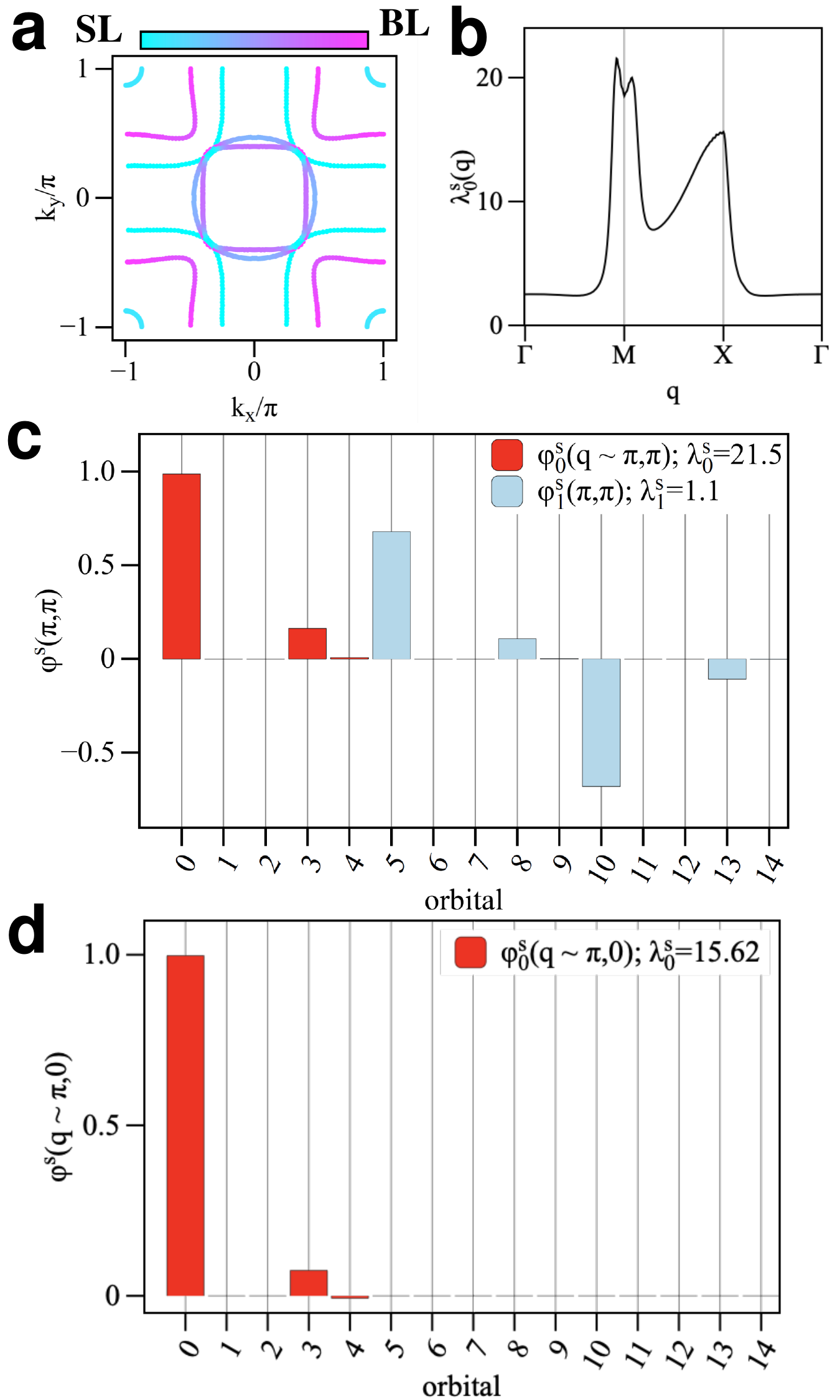}
\caption{ {\bf Magnetic correlations.}  {\bf b} The dominant character of the Fermi-surface Bloch states arising from the SL (magenta) and BL (cyan) sublattices.
{\bf b} Leading eigenvalue $\lambda^s_0({\bf q})$ of the spin susceptibility tensor $\chi^s_{\ell_1\ell_1\ell_2\ell_2}({\bf q})$, where $\ell_i$ are the fifteen Ni-$d$ orbitals included in the model, along a high-symmetry path in the BZ. {\bf c} Leading and sub-leading eigenvectors $\varphi^s_\alpha(\pi,\pi)$ of $\chi^s_{\ell_1\ell_1\ell_2\ell_2}({\bf q})$ for its maximum at ${\bf q}=(\pi,\pi)$ with corresponding eigenvalues $\lambda^s_0=21.5$ and $\lambda^s_1=1.1$, respectively.
{\bf c} The eigenvectors $\varphi^s_\alpha(\pi, 0)$ of $\chi^s_{\ell_1\ell_1\ell_2\ell_2}({\bf q})$ for its maximum at ${\bf q}=(\pi,0)$ with corresponding eigenvalues $\lambda^s_0=15.62$. The orbital number labels 0-4, 5-9, and 10-14 are representing $d_{3z^2-r^2}$, $d_{xz}$, $d_{yz}$, $d_{x^2-y^2}$, and $d_{xy}$ orbitals for the three Ni sites (Ni1 to Ni3 as marked in Fig.~\ref{crystal}~{\bf d}). Here we used Coulomb parameters $U=0.55$ eV, $U'=U/2$, and $J=J'=U/4$, and the calculation was performed for a
temperature $T=0.02$ eV.}
\label{Spin-susceptibility}
\end{figure}

We note that the pairing strength $\lambda = 0.1878$ for the leading $d_{x^2-y^2}$-wave state for $U=0.55$ eV and $T=0.02$ eV is much stronger than the pairing strength we obtained for the pure BL compound La$_3$Ni$_2$O$_7$ \cite{Zhang:nc24}. This indicates that the system, at the level of the RPA treatment, is much closer to a spin-density wave instability. However, since the pairing is weak in the BL sublattice [magenta regions in Fig.~\ref{Spin-susceptibility}~{\bf a}], which separates the SL blocks where the pairing is strong [cyan regions in Fig.~\ref{Spin-susceptibility}~{\bf a}], we expect that the magnitude of the overall $T_c$ of the system does not necessarly reflect the strong pairing in the SL sublattice alone.

To understand the origin of the leading $d_{x^2-y^2}$ pairing instability, we examine the structure of the RPA spin susceptibility tensor $\chi({\bf q})$, which is obtained from the Lindhart function tensor $\chi_0({\bf q})$ as
\begin{eqnarray}
\chi({\bf q}) = \chi_0({\bf q})[1-{\cal U}\chi_0({\bf q})]^{-1}.
\end{eqnarray}
Here, all the quantities are rank-four tensors in the orbital indices $\ell_1, \ell_2, \ell_3, \ell_4$ and ${\cal U}$ is a tensor involving the interaction parameters~\cite{Graser2009}.
The physical spin susceptibility is obtained by summing the pairwise diagonal parts of the tensor, $\chi_{\ell_1\ell_1\ell_2\ell_2}({\bf q})$, over $\ell_1$, $\ell_2$.

As shown in Fig.~\ref{Spin-susceptibility}~{\bf b}, the leading eigenvalue $\lambda^s_0({\bf q})$ of $\chi_{\ell_1\ell_1\ell_2\ell_2}({\bf q})$ shows a strong peak at ${\bf q}=(\pi, \pi)$ (near the M-point) and another peak at ${\bf q}=(\pi, 0)$ (X-point). These peaks correspond to scattering with  $q^M=(\pi,\pi)$ between the $\alpha_1$ and the $\gamma_1$ sheets, and with $q^X=(\pi, 0)$ between the $\gamma_1$ sheets, as shown in Fig.~\ref{RPA}~{\bf a} and {\bf b}.

By examining the corresponding leading and subleading eigenvectors of the susceptibility matrix, $\varphi^s_0({\bf q})$ and $\varphi^s_1({\bf q})$, respectively, for the peak at ${\bf q}=(\pi,\pi)$, we find that the main contribution to this peak comes from the $e_g$ orbitals and the contribution from $t_{2g}$ orbitals is negligible, as displayed in Fig.~\ref{Spin-susceptibility}~{\bf c}.

Furthermore, we see that the dominant scattering near $(\pi, \pi)$ with eigenvalue $\lambda^s_0=21.5$ arises from the SL sublattice (first five $d$ orbitals). The contribution from the BL sublattice as seen in the subleading eigenvector $\varphi_1^s$ is much weaker with $\lambda^s_1=1.1$. In addition, the structure of $\varphi_1^s$ also shows that the magnetic correlations are antiferromagnetic between the top and bottom layer in the BL sublattice as seen from the opposite sign between the Ni2 and Ni3 sites.
We also plot the eigenvector $\varphi^s_0({\bf q})$ for the peak at ${\bf q}=(\pi, 0)$ [or $(0, \pi)$] in Fig.~\ref{Spin-susceptibility}~{\bf d}. Just like the peak at $(\pi, \pi)$, this peak at the $X$ point also arises from the SL lattice. This stripe state with ferromagnetic tendency along one in-plane direction and antiferromagnetic tendency along another in-plane direction can be understood by the competition between intraorbital and interorbital hopping, as discussed in previous studies~\cite{Lin:prl21,Lin:cp,Lin:prb}.

Our previous RPA calculations found the peak in the magnetic susceptibility was near $q = (\pi, 0)$ or $(0, \pi)$ rather than at $(\pi,\pi)$ in the pure BL stacking system~\cite{Zhang:nc24}. Compared to the pure BL stacking case with orbital site occupations of 0.904 for $d_{3z^2-r^2}$ and 0.596 for $d_{x^2-y^2}$, the electronic densities are 0.989 and 0.635 per site for the $d_{3z^2-r^2}$ and $d_{x^2-y^2}$ orbitals in this ``electron-doped'' BL sublattice of this SL-BL stacking 1212-LNO, respectively. Thus, it is reasonable to expect that the ferromagnetic tendency is reduced, leading to antiferromagnetism in the plane.

\noindent {\small \bf \\Comparison with other hybrid stacking nickelate superlattices\\}
Lastly, we examine other hybrid staking nickelate superlattices with the alternating ABAB superlattice geometry,
where A and B can be SL, BL, and TL nichkelates.
Because of the different nominal Ni $d$ electronic configurations, $d^8$ for  SL La$_2$NiO$_4$, $d^{7.5}$ for BL La$_3$Ni$_2$O$_7$, and $d^{7.333}$ for TL La$_4$Ni$3$O$_10$,
one could anticipate different charge transfer between constituent blocks depending on the combination of these RP nickelates.

\begin{figure*}
\centering
\includegraphics[width=0.92\textwidth]{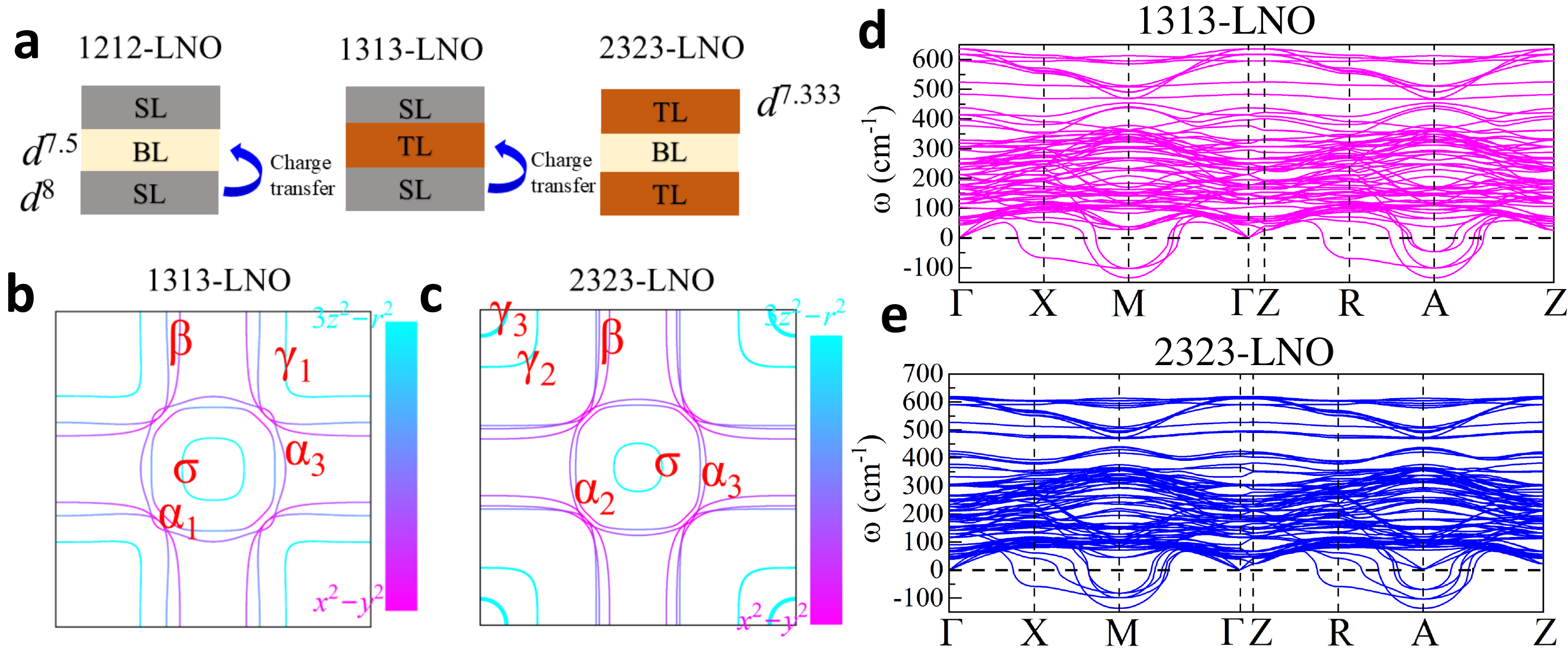}
\caption{ {\bf Comparison with other hybrid stacking RP nickelates.} {\bf a} Sketches of different RP superlattices. Fermi surfaces of P4/mmm phases of {\bf b} 1313-LNO and {\bf c} 2323-LNO under pressure, where the results are obtained based on the optimized crystal structure under pressure from previous works~\cite{Zhang:1313,Zhang:2323}. {\bf d-e} Phonon spectrum of the P4/mmm phase of {\bf d} 1313-LNO and {\bf e} 2323-LNO at 0 GPa, respectively. The coordinates of the high-symmetry points in the Brillouin zone are $\Gamma$ = (0, 0, 0), X = (0, 0.5, 0), M = (0.5, 0.5, 0), Z = (0, 0, 0.5), R = (0, 0.5, 0.5) and A = (0.5, 0.5, 0.5).}
\label{RP-superlattice}
\end{figure*}

There are three different configurations [see Fig.~\ref{RP-superlattice}~{\bf a}]: 1212-LNO, 1313-LNO, and La$_3$Ni$_2$O$_7$/La$_4$Ni$_3$O$_{10}$ (2323-LNO).
In the 1313-LNO, the charge transfer also appears between the SL and the TL layer, leading to hole doping in the SL blocks of 1313-LNO~\cite{Zhang:1313}.
However, in 2323-LNO, there is no obvious charge transfer between BL and TL blocks~\cite{Zhang:2323}.

Based on the optimized crystal structure under pressure from previous works~\cite{Zhang:1313,Zhang:2323}, we computed the Fermi surfaces of 1313-LNO and 2323-LNO in the P4/mmm phase as plotted in Figs.~\ref{RP-superlattice}~{\bf b} and {\bf c}. These Fermi surfaces are also made of $e_g$ states. In the 1313-LNO case, the Fermi surfaces ($\alpha_1$ and $\gamma_1$ sheets) from the SL sublattice look similar to the SL's contribution in the Fermi surface of the 1212-LNO system (see  Fig.~\ref{TB-20}~{\bf d}), but with slightly different shapes. Furthermore, the BL Fermi surfaces ($\beta$ and $\alpha_2$ sheets and $\gamma_2$ pocket) of 1212-LNO are also similar to that in the 2323-LNO system but also have slightly different shapes. This suggests the electrons would slightly self-adjust in different stacking nickelates, leading to small change of shapes of the Fermi surface.

Moreover, we also calculated the phonon spectra of the high-symmetry P4/mmm phase (No.123) without any tilting for the 1313-LNO and 2323-LNO at 0 GPa, by using the density functional perturbation theory approach~\cite{Baroni:Prl,Gonze:Pra1,Gonze:Pra2} analyzed by the PHONONPY software~\cite{Chaput:prb,Togo:sm}. Similarly to 1212-LNO (Fig.~\ref{crystal}~{\bf d}), the phonon dispersion spectra have some imaginary frequencies showing at high symmetry M and A points at 0 GPa for both 1313-LNO and 2323-LNO structures, suggesting the in-plane NiO$_6$ octahedron distortion, leading to a nonflat 180$^{\circ}$ O-Ni-O bond in BL or TL blocks in all three cases. Those also suggest that the tilting distortion can not be fully suppressed by hybrid stacking in nickelate superlattice geometries at ambient conditions, sharing the common structural behavior with other nickelate without pressure.

\noindent {\bf \\Discussion\\}
The very recent efforts studying superconductivity in the RP hybrid superlattice 1212-LNO under pressure extends the study of RP nickelate superconductors in a new direction and provides a new platform to understand superconductivity in the nickelate family~\cite{Nomura:rpp}. Similar to RP BL La$_3$Ni$_2$O$_7$~\cite{Sun:arxiv} and TL La$_4$Ni$_3$O$_{10}$~\cite{Zhu:arxiv11,Li:cpl} nickelate bulk systems, 1212-LNO also shows a superconducting region for a broad pressure range from $\sim 12$ GPa to 25 GPa, the maximum region that was studied, different from the previously reported high-pressure phase diagram of infinite layer nickelates and cuprates superconductors. This suggests that the RP nickelate system is unique and interesting.

Combining first-principles DFT and many-body RPA methods, we have comprehensively studied the 1212-LNO system under pressure (0 GPa to 40 GPa). Without pressure, our phonon calculation indicates that the high-symmetry undistorted P4/mmm phase (No.123) is not stable due to some imaginary frequencies appearing along high symmetry paths. Based on group analysis, three distortion modes $A^ {5+}$, $M^{5+}$, and $M^{5-}$ were obtained, corresponding to Fmmm (No.69), Cmme (No.67), and P2/m (No.10) structures, where the NiO$_6$ octahedron is distorted, leading to the non-flat O-Ni-O bond in the 1212-LNO. Among those structural candidates of 1212-LNO, the Fmmm phase has the lowest energy, where a similar Fmmm structure was proposed to be the solution for 1313-LNO without pressure in experiment~\cite{Puphal:arxiv12}. This suggests that the Fmmm symmetry may be a possible structure for 1212-LNO without pressure. Moreover, our DFT calculations suggest that the Cmmm and P4/mmm phase have almost identical electronic structures due to a very tiny distortion from the P4/mmm to the Cmmm symmetry ($\sim 0.01$ \AA). In this case, our present theoretical results indicate that the Cmmm structure is not the solution of the structure of 1212-LNO without pressure (see more discussion in Supplemental note V), where the NiO$_6$ octahedron is distorted in the plane.  By applying pressure, the P4/mmmm phase becomes stable and all unstable distortion modes were gradually suppressed, leading to the P4/mmm structure at around 12 GPa, where this value is independent of $U_{\rm eff}$. Furthermore, this value of pressure is very close to the critical pressure of the observed superconductivity in 1212-LNO~\cite{Shi:arxiv25}. Thus, it is very interesting to explore the structural stability of the Fmmm phase of 1212-LNO. However, due to limitations of our computing resources, we can not study the phonon spectrum for this Fmmm phase of 1212-LNO, which is a $2\times2\times2\times$ supercell compared to the conventional cell of the P4/mmm structure. Thus, we leave this work to future experimental and theoretical studies for full confirmation.

The electronic structure of 1212-LNO (La$_5$Ni$_3$O$_{11}$) arises from the combination of La$_2$NiO$_4$ and  La$_3$Ni$_2$O$_7$. In addition, we find an obvious ``charge transfer'' effect from the SL to BL sublattices in 1212-LNO, leading to hole-doped SL blocks, compared to La$_2$NiO$_4$ with SL stacking itself. Furthermore, our RPA calculations find a leading $d_{x^2-y^2}$-wave pairing state, followed by three subleading states ($g$-wave, $d_{xy}$-wave, and $s_{\pm}$-wave). The pairing gap is strong for states on the Fermi surface that arise from the SL sublattice and weaker on the outer $\alpha_2$ sheet from the BL sublattice, while the $\beta$ and $\gamma_2$ sheets from the BL blocks do not contribute to the pairing states in the 1212-LNO case. This suggests that pairing is strong in the SL but much weaker than in the BL. In this case, although the calculated pairing strength $\lambda$ is much larger than that previously calculated for the pure BL compound La$_3$Ni$_2$O$_7$, $T_c$ is not necessarily expected to follow this trend since the two SL blocks are separated by a BL block in which the pairing is weak.

In addition, our spin susceptibility calculations demonstrates a strong peak at ${\bf q}=(\pi, \pi)$ and a second peak at ${\bf q}=(\pi, 0)$, corresponding to N\'eel and stripe antiferromagnetic states, respectively. These are caused by the nesting  between the $\alpha_1$ and the $\gamma_1$ sheets and between the $\gamma_1$ sheets, respectively. Our analysis also unveils similarities and differences between this 2323-LNO and other hybrid stacking nickelate superlattice structures. Our results  suggest that the interface could play a possible role in superconductivity by adjusting the charge densities in the nickelate blocks. This picture is quite similar to that recently observed in BL nickelate thin-films on a LaSrAlO$_4$ substrate where the interface induces a hole-doping effect in the BL nickelate sides~\cite{Ko:nature,Zhou:nature}. Thus,
another direction for the study of superconductivity in nickelates at ambient pressure could be to work on interfaces between nickelate blocks and other layer systems to adjust the electronic densities on the nickelate sides.

\noindent {\bf \\Methods\\}

\noindent {\small \bf DFT method\\}
In this work, we employ first-principles DFT calculations by using the Vienna {\it ab initio} simulation package (VASP) software with the projector augmented wave (PAW) method~\cite{Kresse:Prb,Kresse:Prb96,Blochl:Prb}. Furthermore, the electronic correlations
dealt with the generalized gradient approximation (GGA) and the Perdew-Burke-Ernzerhof (PBE) exchange potential~\cite{Perdew:Prl}. Both atomic positions and crystal lattice constants were fully relaxed until the Hellman-Feynman force on each atom was smaller
than $0.01$ eV/{\AA} for various pressures with different structural symmetries. Here, the plane-wave cutoff energy was set as $550$~eV and the k-point mesh was appropriately modified for different structures to make the k-point densities approximately the same in
reciprocal space, such as $20\times20\times5$ for the P4/mmm (No. 123) phase of 1212-LNO. To obtain the hopping matrices and crystal-field splittings, we used the maximally localized Wannier functions (MLWFs) method to fit the Ni's $3d$ bands with DFT bands,
by using the WANNIER90 packages~\cite{Mostofi:cpc}. For the phonon spectrum of the P4/mmm (No. 123) phases of the alternating SL-BL stacking 1212-LNO, a  $2\times2\times2$ supercell structure was used in the phonon calculation, by using the density functional
perturbation theory approach~\cite{Baroni:Prl,Gonze:Pra1}, analyzed by the PHONONPY software in the primitive unit cell~\cite{Chaput:prb,Togo:sm}. All the crystal structures were visualized with the VESTA code~\cite{Momma:vesta}.

\noindent {\small \bf Tight-binding method\\}
Based on the hopping matrices and crystal-field splittings for the P4/mmm phase obtained from MLWFs, we constructed a five-band low-energy $3d$-orbital tight-binding model for SL BL stacking 1212-LNO, where the overall filling was considered as $N = 23.0$,
corresponding to $7.666$ electrons per Ni site (the detailed files of hopping matrices and crystal-field splittings can be found in a separate attachment in the Supplemental Materials).  Based on the obtained Fermi energy, a $4001\times4001$ $k$-mesh was used to calculate the Fermi surface.
Although the Fermi surface is made of two $e_g$ orbitals, just using a two-orbital model, we could not perfectly reproduce the DFT band structure then we used the five-orbital in our further calculations (details are shown in Supplemental note V).

\noindent {\small \bf RPA method\\}
To study magnetic correlations and possible superconducting pairing properties, we used the many-body RPA method here, based on a perturbative weak-coupling expansion in the Hubbard interaction, similar to our previous works.
The full Hubbard model Hamiltonian for the BL plus SL discussed here, includes the kinetic energy and interaction terms, and it is written as $H = H_{\rm k} + H_{\rm int}$.
The electronic interaction portion of the Hamiltonian includes the standard intraorbital Hubbard repulsion $U$, the interorbital Hubbard repulsion $U'$, the Hund's coupling $J$, and the on-site interorbital electron-pair hopping term $J'$. Formally, it is given by:
\begin{eqnarray}
H_{\rm int}= U\sum_{i\gamma}n_{i \gamma\uparrow} n_{i \gamma\downarrow} +(U'-\frac{J}{2})\sum_{\substack{i\\\gamma < \gamma'}} n_{i \gamma} n_{i\gamma'} \nonumber \\
-2J \sum_{\substack{i\\\gamma < \gamma'}} {{\bf S}_{i \gamma}}\cdot{{\bf S}_{i \gamma'}}+J \sum_{\substack{i\\\gamma < \gamma'}} (P^{\dagger}_{i\gamma} P^{\phantom{\dagger}}_{i\gamma'}+H.c.),
\end{eqnarray}
where the standard relations $U'=U-2J$ and $J' = J$ are assumed, with $n_{i \gamma} = n_{i \gamma \uparrow}+ n_{i \gamma \downarrow}$ and $P_{i\gamma}$=$c_{i\gamma\downarrow} c_{i\gamma \uparrow}$.

In the multi-orbital RPA approach \cite{Kubo2007,Graser2009,Altmeyer2016,Romer2020}, the enhanced spin susceptibility is obtained from the bare susceptibility (Lindhart function) via $\chi_0({\bf q})$ as $\chi({\bf q}) = \chi_0({\bf q})[1-{\cal U}\chi_0({\bf q})]^{-1}$. Here, $\chi_0({\bf q})$ is an orbital-dependent susceptibility tensor and ${\cal U}$ is a tensor that contains the intra-orbital $U$ and inter-orbital $U'$ density-density interactions, the Hund's rule coupling $J$, and the pair-hopping $J'$ term. The pairing strength $\lambda_\alpha$ for channel $\alpha$ and the corresponding gap structure $g_\alpha({\bf k})$ are obtained from solving an eigenvalue problem of the form
\begin{eqnarray}\label{eq:pp}
	\int_{FS} d{\bf k'} \, \Gamma({\bf k -  k'}) g_\alpha({\bf k'}) = \lambda_\alpha g_\alpha({\bf k})\,,
\end{eqnarray}
where the momenta ${\bf k}$ and ${\bf k'}$ are on the FS and $\Gamma({\bf k - k'})$ contains the irreducible particle-particle vertex. In the RPA approximation, the dominant term entering $\Gamma({\bf k-k'})$ is the RPA spin susceptibility $\chi({\bf k-k'})$.

\noindent {\bf {\small \\ Data availability\\}} The input parameters for our TB calculations are available in a sperate file in the Supplementary information. Any additional data that support the findings of this study are available from the corresponding author upon request.
\noindent {\bf {\small \\ Code availability\\}} The {\it ab initio} calculations were done with the VASP code, which is available under commercial license at \href{https://www.vasp.at/}{www.vasp.at}.  Simulations of our home-made RPA codes are available from the corresponding authors
upon reasonable request.

\noindent {\bf {\\Acknowledgements\\}}
\small {The work was supported by the U.S. Department of Energy (DOE), Office of Science, Basic Energy Sciences (BES), Materials Sciences and Engineering Division.}

\noindent {\bf {\\ Author contributions\\}} \small {Y.Z., L.-F.L., T.A.M., and E.D. designed the project. Y.Z., L.F.L., and T.A.M. carried out numerical calculations for DFT, the TB model, and RPA calculations.  All co-authors provided useful comments and discussion on the paper and  wrote the manuscript.}

\noindent {\bf {\\Competing interests\\}} \small {The authors declare no competing interest.}

\noindent {\bf {\\ Additional information\\}}
Correspondence should be addressed to Ling-Fang Lin, Thomas A. Maier or Elbio Dagotto.

\noindent \small {{\bf Supplementary information} The online version contains supplementary material available at \href{xx/xxxxxx}{xx/xxxxxx.}}

\end{document}